\documentclass[aps,prl,reprint,superscriptaddress,10pt]{revtex4-1}
\usepackage{amsmath}
\usepackage{hyperref}
\usepackage{graphicx}
\usepackage{epstopdf}

\newcommand{\primedvector}[1]{\vec{#1}\mkern2mu\vphantom{#1}}

\begin{document}

\title{Highly-polarizable ion in a Paul trap}

\author{Gerard Higgins}
\email[]{gerard.higgins@fysik.su.se}
\affiliation{Department of Physics, Stockholm University, SE-106 91 Stockholm,
Sweden}
\affiliation{Institut f\"ur Experimentalphysik, Universit\"at Innsbruck, 
AT-6020 Innsbruck, Austria}
\author{Fabian Pokorny}
\affiliation{Department of Physics, Stockholm University, SE-106 91 Stockholm,
Sweden}
\author{Chi Zhang}
\affiliation{Department of Physics, Stockholm University, SE-106 91 Stockholm,
Sweden}
\author{Markus Hennrich}
\affiliation{Department of Physics, Stockholm University, SE-106 91 Stockholm,
Sweden}

\date{\today}

\begin{abstract}
Usually the influence of the quadratic Stark effect on an ion's trapping potential is minuscule and only needs to be considered in atomic clock experiments.
In this work we excite a trapped ion to a Rydberg state with polarizability $\sim$~eight orders of magnitude higher than a low-lying electronic state; we find that the highly-polarizable ion experiences a vastly different trapping potential owing to the Stark effect.
We observe changes in trap stiffness, equilibrium position and minimum potential, which can be tuned using the trapping electric fields.
These effects lie at the heart of proposals to shape motional mode spectra, simulate quantum magnetism and coherently drive structural phase transitions; in addition we propose using these effects to simulate cosmological particle creation, study quantum fluctuations of work and minimize ion micromotion.
Mitigation of Stark effects is important for coherent control of Rydberg ions; we illustrate this by carrying out the first Rabi oscillations between a low-lying electronic state and a Rydberg state of an ion.
\end{abstract}

\maketitle
Although ions in Paul traps are trapped close to electric field nulls, they experience non-zero electric fields which modify the trapping potential via the quadratic Stark effect $\Delta E = -\frac{1}{2} \alpha \mathcal{E}^2$.
While this modification is small for atomic ions in low-lying electronic (LLE) states, it strongly affects the trapping potential of highly-polarizable Rydberg ions, causing striking phenomena to emerge when trapped ions are excited to Rydberg states.

Because the trap stiffness is modified by the Stark effect, atomic transition frequencies depend on motional phonon number and differential polarizabilities.
This temperature-dependent Stark shift must be accounted for in trapped ion atomic clocks which operate below the 10\textsuperscript{-17} level \cite{Rosenband2008, Huang2012, Tamm2014}.
We observe this effect directly for the first time by exciting an ion to a highly-polarizable Rydberg state.

We find that a static offset electric field changes the equilibrium position and the minimum potential of the trapping potential of a highly-polarizable Rydberg ion relative to that of an ion in a LLE state.
The change in minimum potential alters the Rydberg-excitation energy; this shift may be used to reduce micromotion beyond the state-of-the-art level.
Micromotion minimalization is critical for trapped ion atomic clocks \cite{Berkeland1998, Keller2015} and for studies of atom-ion collisions in the quantum regime \cite{Grier2009, Schmid2010, Zipkes2010}.
The change in trap position allows us to strongly drive phonon-number changing transitions during Rydberg excitation.
The shifted trapping potential means our system may be used for studying quantum fluctuations of work \cite{Huber2008, An2014} or for simulations of molecular dynamics \cite{Li2012}.

Trapped Rydberg ions are an exciting new platform for quantum information processing \cite{Mueller2008}, which combines the exquisite control of trapped ion systems with the strong interactions of Rydberg atoms.
Stark effects must be mitigated for Rydberg ions to be coherently controlled; we illustrate this by driving the first Rabi oscillations between a LLE state and a Rydberg state of a trapped ion.

This study provides a solid foundation for further work which utilizes the change of trap stiffness during Rydberg excitation, including vibrational mode shaping in large ion strings \cite{Li2013}, coherent driving of structural phase transitions \cite{Li2012} and simulations of quantum magnetism \cite{Nath2015}.
In addition we propose using this effect to simulate cosmological particle creation \cite{Schuetzhold2007, Wittemer2019}.

In a linear Paul trap ions are confined by a combination of oscillating and static electric quadrupole fields.
When the quadrupole field nulls overlap, the electric potential near the center is
\begin{equation*}
\Phi(t) = A \cos{\Omega t}(x^2 - y^2) - B \left((1+\epsilon)x^2 + (1-\epsilon)y^2 - 2z^2\right)
\end{equation*}
where $A$ and $B$ are the electric field gradients of the oscillating and static electric quadrupole fields, $\Omega$ is the frequency of the oscillating field, and $\epsilon$ accounts for non-degeneracy of the radial trapping frequencies.

Although ions are dynamically trapped, the effective trapping potential can be described by the time-independent harmonic pseudopotential \cite{Major2005}
\begin{equation*}
U = \tfrac{1}{2} M \left( \omega_x^2 x^2 + \omega_y^2 y^2 + \omega_z^2 z^2 \right)
\end{equation*}
where $M$ is the ion mass and the trapping frequencies $\omega_{x,y,z}$ depend on $A$, $B$, $\epsilon$ and $\Omega$ as described in \cite{appendix}.

A trapped ion with polarizability $\alpha$ experiences Stark shift $\Delta U = - \tfrac{1}{2} \alpha \mathcal{E}(t)^2$, where the electric field $\vec{\mathcal{E}}(t) = -\vec{\nabla}{\Phi(t)}$.
When considering timescales much longer than the trap drive period $\frac{2 \pi}{\Omega}$ the squared electric field strength can be time-averaged and $\Delta U$ takes the form of an additional harmonic potential
\begin{align} \label{eq_additional_harmonic_potential}
\begin{split}
\Delta U &= - \tfrac{1}{2} \alpha \langle \mathcal{E}(t)^2 \rangle
\\
&\approx - \alpha A^2 \left( x^2 + y^2 \right)
\end{split}
\end{align}
where the approximation uses $A^2 \gg B^2$, which is usually satisfied.
A polarizable ion experiences altered trapping potential $U' = U + \Delta U$ with altered trapping frequencies
\begin{equation} \label{eq_omega_prime}
\omega_{x}' \approx \sqrt{\omega_{x}^2 - \frac{2 \alpha A^2}{M}},\:\: \omega_{y}' \approx \sqrt{\omega_{y}^2 - \frac{2 \alpha A^2}{M}},\:\: \omega_{z}' \approx \omega_{z}
\end{equation}

\begin{figure}[ht]
\centering
\includegraphics[width=\columnwidth]{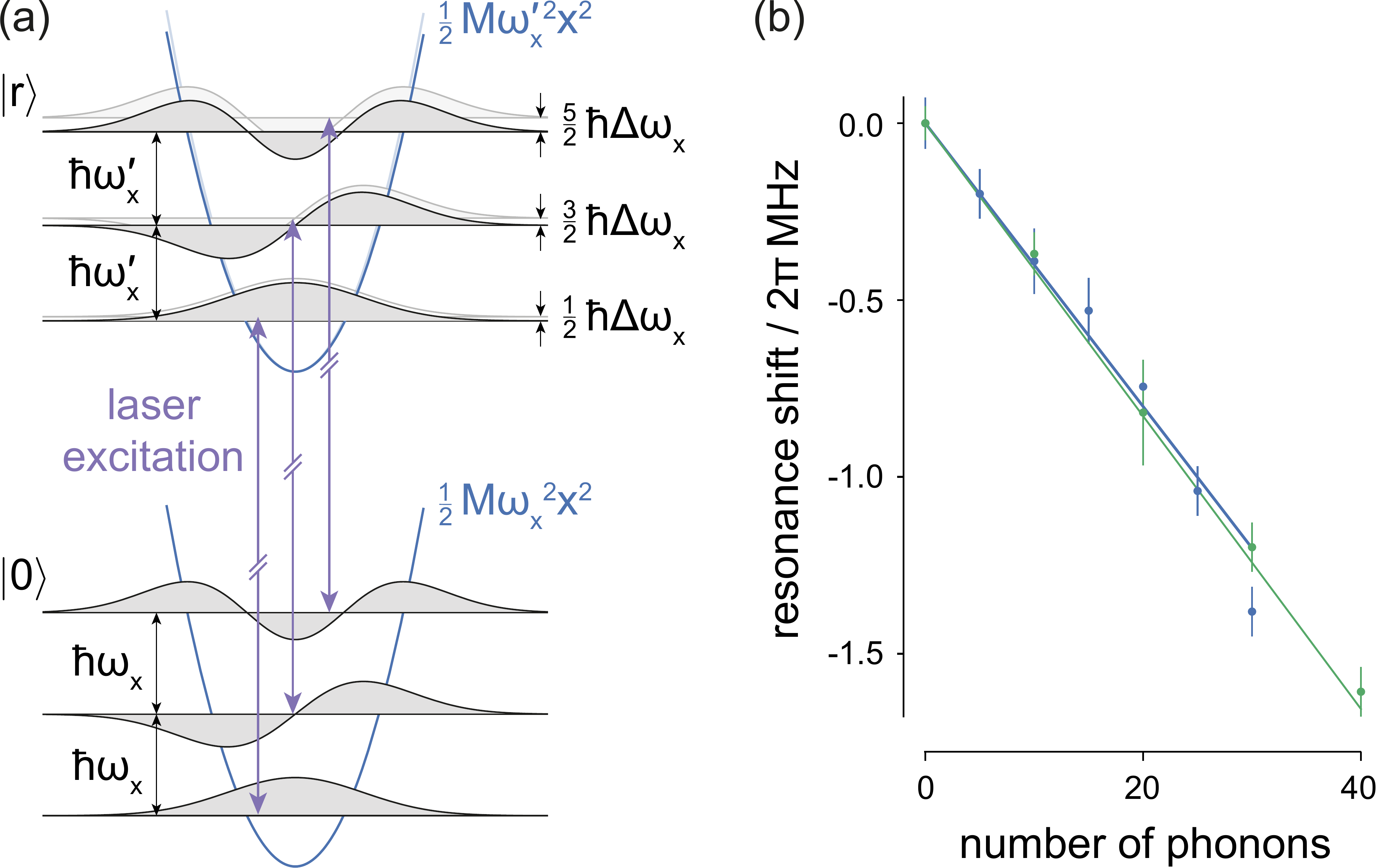}
\caption{Transition energy between two states with different polarizabilities depends on phonon number.
(a) Illustration of the trapping potential and phonon states in the x-direction for an ion in a LLE state with negligible polarizability $|0\rangle$ and a highly-polarizable Rydberg state $|r\rangle$; $\Delta \omega_x = \omega_x' - \omega_x$.
(b) The $4D_{5/2}\leftrightarrow46S_{1/2}$ resonance frequency decreases as the number of phonons in the x-mode (y-mode) is increased while the number of phonons in the y-mode (x-mode) is fixed at $0$; data points are blue (green).
The theory lines use Eq.~(\ref{eq_phonon_number_dependent_energy_shift}) with $\omega_x'-\omega_x=-2\pi\times 40.1\,\mathrm{kHz}$ and $\omega_y'-\omega_y=-2\pi\times 41.4\,\mathrm{kHz}$.
Error bars (68\% confidence interval) are dominated by frequency drifts of the Rydberg-excitation lasers.
}
\label{fig1}
\end{figure}

As a result, the transition frequency between two atomic states with polarizabilities $\{\alpha_1, \alpha_2\}$ depends on the number of phonons in radial motional modes $\{n_x, n_y\}$,
\begin{align}
\begin{split} \label{eq_phonon_number_dependent_energy_shift}
\Delta E_{1 \rightarrow 2} = &\left(n_x+\tfrac{1}{2}\right) \hbar\big(\omega_x'(\alpha_2)-\omega_x'(\alpha_1)\big) \\
+ &\left(n_y+\tfrac{1}{2}\right) \hbar\big(\omega_y'(\alpha_2)-\omega_y'(\alpha_1)\big)
\end{split}
\end{align}
as illustrated in Fig.~\ref{fig1}(a).
We confirm Eq.~(\ref{eq_phonon_number_dependent_energy_shift}) experimentally as follows:
We prepare a single trapped $\mathrm{^{88}Sr^+}$ ion in number states of the radial modes and measure the dependence of the $4D_{5/2} \leftrightarrow 46S_{1/2}$ transition frequency on the radial phonon number. Preparation of phonon number states and detection of Rydberg excitation is described in \cite{appendix}. Results are shown in Fig.~\ref{fig1}(b).
The model line uses Eq.~(\ref{eq_phonon_number_dependent_energy_shift}) and the theory value for polarizability $\alpha_{46S}=5.6\times10^{-31}\,\mathrm{Cm^2V^{-1}}$.
The polarizability of $4D_{5/2}$ is eight orders of magnitude smaller than $\alpha_{46S}$ \cite{Jiang2009} and is negligible in our experiments.
The energy shift in Eq.~(\ref{eq_phonon_number_dependent_energy_shift}) results from the mean-squared electric field strength sampled by the polarizable ion during secular motion; inclusion of intrinsic micromotion changes the model prediction by $\approx 2 \%$ \cite{appendix}.

Because the transition frequency between states with different polarizabilities depends on phonon number, an ion with a broad radial phonon distribution will display a broader resonance linewidth than an ion with a narrow radial phonon distribution.
As a result narrow radial phonon distributions and thus near-ground-state cooling is advantageous for trapping ion atomic clocks which operate below the 10\textsuperscript{-17} level.
This effect was used to explain differences between the Rydberg-excitation spectral lineshapes of a sideband-cooled ion and a Doppler-cooled ion in \cite{Higgins2017a}.

The quadratic Stark effect is manifest in a more striking fashion when the null of the static electric quadrupole field ($\vec{r}_{\mathrm{dc}}$) and the null of the oscillating electric quadrupole field ($\vec{r}_{\mathrm{rf}}$) do not overlap.
In such a setup the ion equilibrium position depends on its polarizability; as a result phonon-number changing transitions may be driven strongly between states with different polarizabilities \cite{Mueller2008}.

When a static electric field
\begin{equation*} \label{eq_d_offset_electric_field}
\vec{\mathcal{E}}_{\mathrm{offset}}=-2B\left((1+\epsilon)x_{\mathrm{dc}}, (1-\epsilon)y_{\mathrm{dc}}, 0\right)
\end{equation*}
is applied to the system the static quadrupole field null is shifted to $\vec{r}_{\mathrm{dc}} = (x_{\mathrm{dc}},y_{\mathrm{dc}},0)$ while the oscillating quadrupole field null is unchanged $\vec{r}_{\mathrm{rf}} = (0,0,z)$.
The electric potential becomes
\begin{align}
\Phi(&t) = A \cos{\Omega t}(x^2 - y^2) \label{eq_electric_pot_w_mm} \\
&- B ((1+\epsilon)(x-x_{\mathrm{dc}})^2 + (1-\epsilon)(y-y_{\mathrm{dc}})^2 - 2z^2) \notag 
\end{align}
and the harmonic pseudopotential becomes \cite{Berkeland1998}
\begin{equation*}
U = \tfrac{1}{2} M \left( \omega_x^2 (x-x_{\mathrm{eq}})^2 + \omega_y^2 (y-y_{\mathrm{eq}})^2 + \omega_z^2 z^2 \right)
\end{equation*}
where
\begin{equation*}
x_{\mathrm{eq}} = - \frac{2 e B (1+\epsilon) x_{\mathrm{dc}}}{M \omega_x^2},\:\: y_{\mathrm{eq}} = - \frac{2 e B (1-\epsilon) y_{\mathrm{dc}}}{M \omega_y^2} \label{eq_equilibria_positions}
\end{equation*}
and the equilibrium position $\vec{r}_{\mathrm{eq}}=(x_{\mathrm{eq}},y_{\mathrm{eq}},0)$.

\begin{figure}[ht]
\centering
\includegraphics[width=\columnwidth]{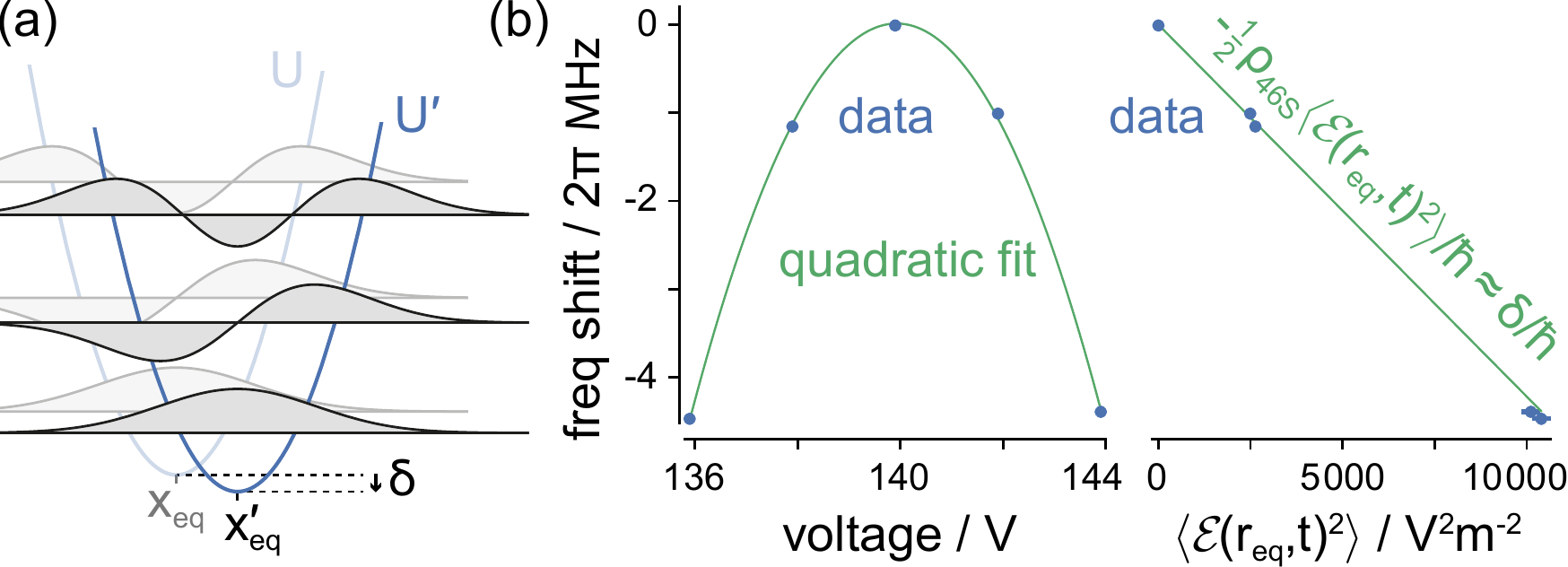}
\caption{The trapping potential of a polarizable ion is shifted when the nulls of the trapping electric quadrupole fields do not overlap.
(a) Schematic, (b) Energy shift $\delta$ is measured: an offset field is varied via the voltage on an electrode and the Rydberg excitation-energy responds quadratically.
The x-axis is rescaled \cite{appendix} to show the shift follows Eq.~(\ref{eq_energy_shift}).
Frequency drifts of the Rydberg-excitation lasers cause errors too small to be visible in this plot, x-error bars from uncertainty in x-axis rescaling are just discernible (68\% confidence interval).
}
\label{fig2}
\end{figure}

A polarizable ion experiences the additional harmonic potential $\Delta U$ from Eq.~(\ref{eq_additional_harmonic_potential}) and $U'$ becomes
\begin{equation*}
U' = \tfrac{1}{2} M \left( {\omega_x'}^2 \left(x-x_{\mathrm{eq}}'\right)^2 + {\omega_y'}^2 \left(y-y_{\mathrm{eq}}'\right)^2 + {\omega_z}^2 z^2 \right) + \delta
\end{equation*}
where the equilibrium position is ${\primedvector{r}'_{\mathrm{eq}}}=(x_{\mathrm{eq}}',y_{\mathrm{eq}}',0)$ with
\begin{equation}
x_{\mathrm{eq}}' = x_{\mathrm{eq}} \left( 1 - \frac{2 \alpha A^2}{M \omega_x^2} \right)^{-1},\:\:
y_{\mathrm{eq}}' = y_{\mathrm{eq}} \left( 1 - \frac{2 \alpha A^2}{M \omega_y^2} \right)^{-1} \label{eq_Ryd_eq_pos}
\end{equation}
and the energy shift
\begin{align} \label{eq_energy_shift}
\begin{split}
\delta &= \tfrac{1}{2} M \left( \omega_x^2 x_{\mathrm{eq}}^2 +\omega_y^2 y_{\mathrm{eq}}^2 - {\omega_x'}^2 {x'_{\mathrm{eq}}}^2 - {\omega_y'}^2 {y'_{\mathrm{eq}}}^2 \right)
\\
&\approx - \alpha A^2 (x_{\mathrm{eq}}^2+y_{\mathrm{eq}}^2)
= - \tfrac{1}{2} \alpha \langle \mathcal{E}\left( \vec{r}_{\mathrm{eq}}, t \right)^2 \rangle
\end{split}
\end{align}
The approximation in Eq.~(\ref{eq_energy_shift}) reveals $\delta$ is a quadratic Stark shift which results from the electric field at the equilibrium position $\vec{r}_{\mathrm{eq}}$ acting on the polarizable ion.
This approximation is valid provided $|\alpha A^2| \ll \frac{1}{2} M \omega_{x,y}^2$ such that the Stark shift $\Delta U$ can be treated as a perturbation.
It is worth noting that although $\delta$ appears when there is excess micromotion in the system it does not result from this motion.
The differences between $U$ and $U'$ are represented in Fig.~\ref{fig2}(a).

We measure $\delta$ in the experiment:
Our setup includes electrodes which we use to control $\vec{\mathcal{E}}_{\mathrm{offset}}$ to overlap the nulls of the quadrupole fields and thus minimize micromotion.
We vary the voltage applied to one of these electrodes and measure a quadratic response of the $4D_{5/2} \leftrightarrow 46S_{1/2}$ resonance frequency, shown in Fig.~\ref{fig2}(b).
The x-axis is rescaled to $\langle \mathcal{E}\left(r_{\mathrm{eq}},t\right)^2\rangle$ \cite{appendix} and we see the frequency shift obeys Eq.~(\ref{eq_energy_shift}).
This shift was observed previously in transitions between LLE states \cite{Yu1994, Schneider2005}; here the shift is several orders of magnitude larger owing to the giant Rydberg ion polarizability.
The additional shift due to the altered trap stiffness in Eq.~(\ref{eq_phonon_number_dependent_energy_shift}) is negligible here since sideband cooling is employed.

Excess micromotion vanishes at the turning point in Fig.~\ref{fig2}(b); Rydberg spectroscopy may thus be used to minimize micromotion.
We use this method to reduce the residual oscillating field strength at the equilibrium position $\vec{r}_{\mathrm{eq}}$ to $\mathcal{E}_{\mathrm{res}}\approx20\,\mathrm{Vm^{-1}}$.
We expect this method will allow us to reduce the residual oscillating field strength beyond the state-of-the-art $\mathcal{E}_{\mathrm{res}}\approx0.3\,\mathrm{Vm^{-1}}$ level \cite{Harter2013,Keller2015} after we improve the frequency stability of the Rydberg-excitation lasers and when we employ Rydberg states with higher polarizabilities \cite{appendix}.
This method has the advantage that it is sensitive to micromotion in all directions with a single probing setup.
Stark shifts of neutral Rydberg atoms have already been used to minimize stray electric fields in neutral atom systems \cite{Osterwalder1999}.

\begin{figure*}[ht!]
\centering
\includegraphics[width=\textwidth]{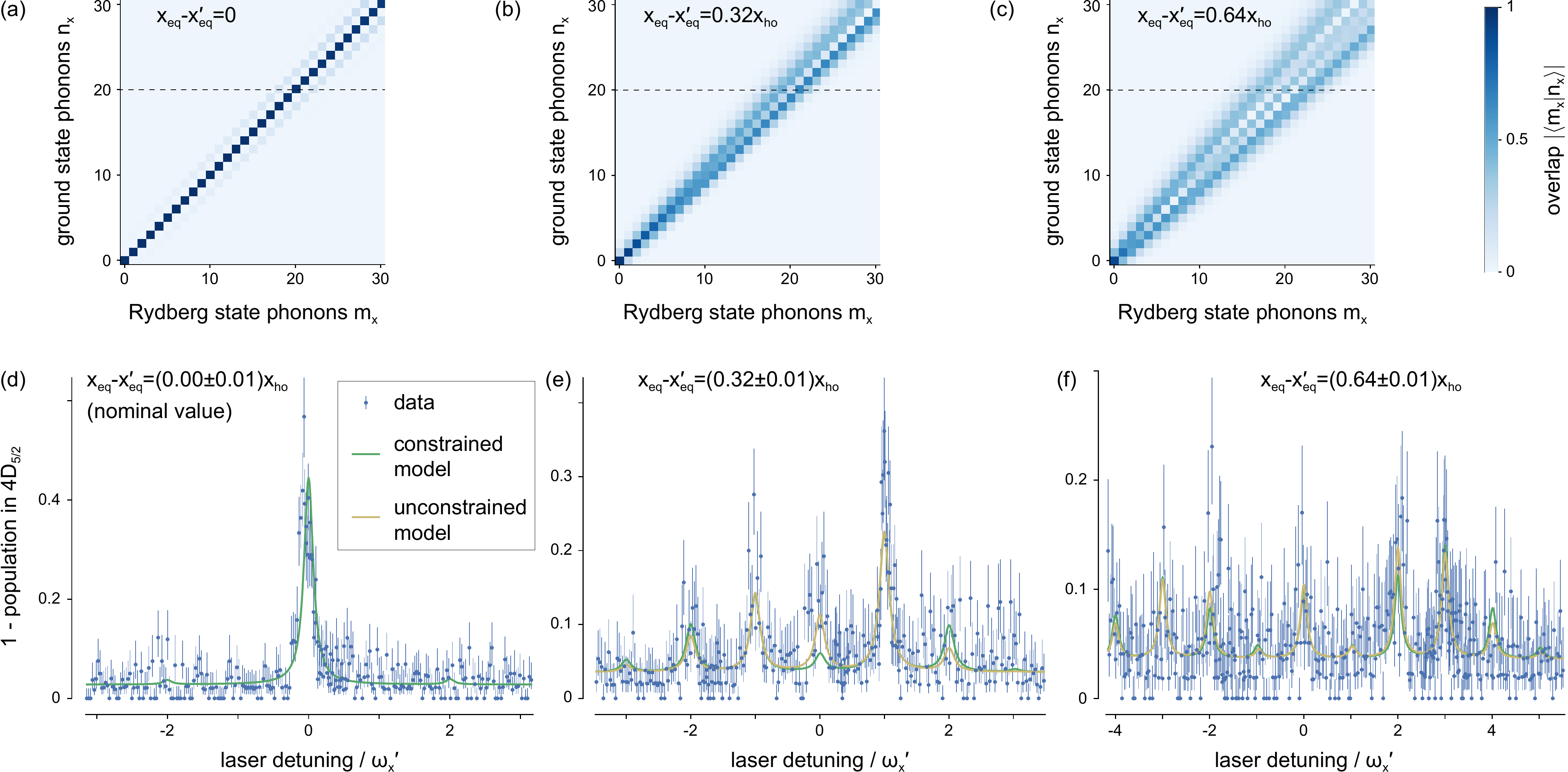}
\caption{Phonon states can be changed during a transition between atomic states with different polarizabilities.
(a--c)~Overlap integrals of phonon number states of an ion in a LLE state with negligible polarizability and phonon states of a highly-polarizable Rydberg ion.
The shift of the one-dimensional harmonic pseudopotentials $x_{\mathrm{eq}}-x'_{\mathrm{eq}}$ increases from (a)$\rightarrow$(c); $x_{\mathrm{ho}}$ is the harmonic oscillator characteristic length.
As $x_{\mathrm{eq}}-x'_{\mathrm{eq}}$ increases and as the phonon number increases, the overlap of phonon states with different quantum numbers also increases and phonon-number changing transitions become easier to drive.
(d--f)~Rydberg-excitation spectra for an ion prepared with $n_x=20$ and $n_y=0$.
Resonances at laser detuning $N\times\omega_x'$ correspond to transitions which increase the phonon number by $N$. 
From (d)$\rightarrow$(f) $x_{\mathrm{eq}}-x'_{\mathrm{eq}}$ is increased (by controlling an offset electric field $\vec{\mathcal{E}}_{\mathrm{offset}}$) and phonon-number changing transitions are driven more strongly.
The amplitudes of the resonances in (d--f) are described by $n_x=20$ slices through (a--c).
The spectra are modeled \cite{appendix} by Lorentzian absorption lines separated by $\omega_x'$ with amplitudes given by the overlap integrals of the phonon states.
The green model curves use only experimentally-constrained parameters while the orange curves treat $x_{\mathrm{eq}}-x'_{\mathrm{eq}}$ as a free parameter and show better agreement with the experimental data.
Error bars indicate projection noise (68\% confidence interval).
}
\label{fig3}
\end{figure*}

Strong phonon-number changing transitions can be driven between states with different polarizabilities:
The motional states of an ion in trapping potential $U$ ($U'$) are solutions of the quantum harmonic oscillator, parameterized by trapping frequencies $\omega_{x,y,z}$ ($\omega_{x,y}'$ and $\omega_z$) and $\vec{r}_{\mathrm{eq}}$ ($\primedvector{r}'_{\mathrm{eq}}$).
Electronic states with different polarizabilities have different trapping potentials and different motional eigenstates.
Overlaps between motional states are called Franck-Condon factors, and are plotted in Fig.~\ref{fig3}(a-c).
The figure shows phonon-number changing transitions are stronger with increasing separation between equilibrium positions and with increasing phonon number.
Phonon-number changing transitions also become stronger with increasing difference between trapping frequencies.
Note that these phonon-number changing transitions result from the Stark effect, whereas phonon-number changing transitions usually present in trapped ion experiments result from the Doppler effect \cite{Wineland1998, appendix} or magnetic field gradients \cite{Ospelkaus2008, Johanning2009, Srinivas2018}.

We probe the strong phonon-number changing transitions as follows: We prepare a single ion in motional state $n_x=20, n_y=0$ and measure Rydberg-excitation spectra as we vary the voltage on a micromotion-minimization electrode -- in this manner we vary $\mathcal{E}_{\mathrm{offset}}$ and thus $|\vec{r}_{\mathrm{eq}}-\primedvector{r}_{\mathrm{eq}}|$.
The results are shown in Fig.~\ref{fig3}(d-f).
The heights of the peaks in Fig.~\ref{fig3}(d-f) are related to $n_x=20$ slices through Fig.~\ref{fig3}(a-c).
Qualitative agreement is observed between experimental data and model curves which use experimentally-constrained parameters \cite{appendix}.
Reasonable quantitative agreement is observed when $|\vec{r}_{\mathrm{eq}}-\primedvector{r}'_{\mathrm{eq}}|$ is treated as a free parameter, this suggests an additional offset electric field was present and not accounted for.

Franck-Condon factors are ubiquitous in transitions between molecular states, and thus trapped Rydberg ions may serve as a natural system for quantum simulation of molecular systems \cite{Li2012}.
The difference between the trapping potentials $U$ and $U'$ may also allow quantum fluctuations of work to be investigated \cite{Huber2008, An2014}.

Because $\mathcal{E}(t)$ oscillates with frequency $\Omega$ the quadratic Stark shift oscillates with frequency $2\Omega$ and sidebands appear in excitation spectra at even multiples of $\Omega$, provided $\alpha \langle\mathcal{E}(t)^2\rangle/2 \not\ll \hbar (2\Omega)$.
These Stark sidebands are distinct from the usual micromotion sidebands described by the first-order Doppler effect \cite{Berkeland1998}.
Stark sidebands are used to describe spectral lineshapes in another Rydberg ion experiment \cite{Feldker2015, Mokhberi2019}.
In our system $\alpha \langle\mathcal{E}^2\rangle/2 \ll \hbar (2\Omega)$ and Stark sidebands are not observed.

Just as with high-precision trapped ion atomic clocks, quantum information processing with Rydberg ions requires mitigation of the quadratic Stark effect, since high-fidelity operations require addressing of individual narrow transition lines.
Stark effects may be mitigated by using microwave-dressed Rydberg states with vanishing polarizabilities \cite{Li2014} (cf.\ ideal clock transitions have small differential polarizabilities \cite{Ludlow2015});
alternatively $\langle \mathcal{E}^2 \rangle$ may be minimized by ground-state cooling of radial motional modes and by overlapping the nulls of the quadrupole fields (i.e.\ micromotion minimization).

\begin{figure}[ht!]
\centering
\includegraphics[width=0.65\columnwidth]{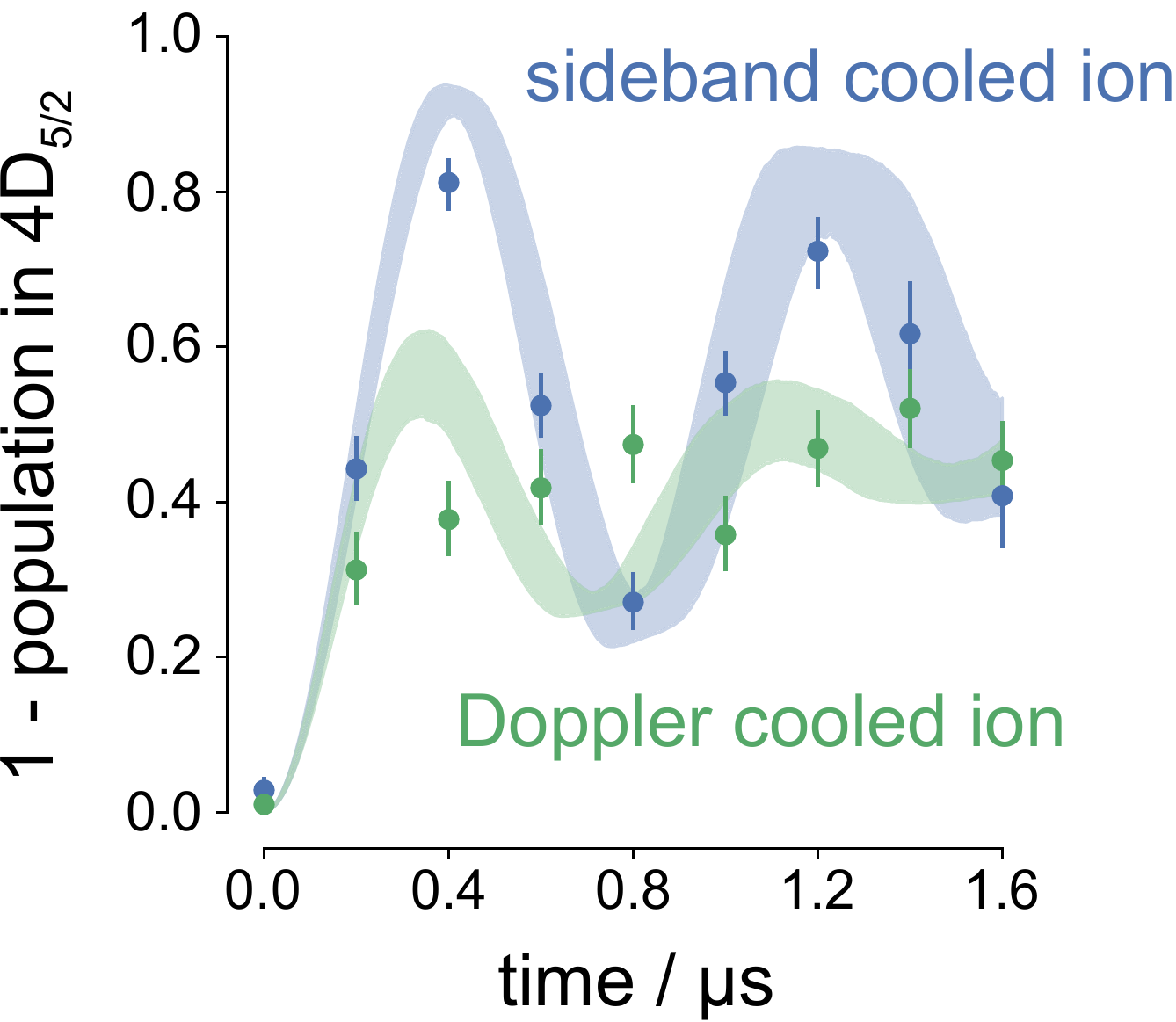}
\caption{Rabi oscillations between LLE state $4D_{5/2}$ and Rydberg state $46S_{1/2}$ are observed when radial sideband cooling is employed, while oscillations are washed out when a warmer Doppler-cooled ion is used.
Error bars indicate projection noise (68\% confidence interval) and shaded areas show the central 68\% of the Monte Carlo simulation results \cite{appendix}.}
\label{fig4}
\end{figure}

We illustrate the importance of mitigating the quadratic Stark effect by comparing Rabi oscillations between LLE state $4D_{5/2}$ and Rydberg state $46S_{1/2}$ for a sideband-cooled ion (with a narrow radial phonon distribution) and a Doppler-cooled ion (with a broad radial phonon distribution) in Fig.~\ref{fig4}: Rabi oscillations are visible for the sideband-cooled ion, while they are smeared out for the Doppler-cooled ion owing to its broader resonance linewidth which stems from Eq.~(\ref{eq_phonon_number_dependent_energy_shift}).
The parameters of the theory curves are experimentally constrained, as described in \cite{appendix}.
Although coherent excitation of a Rydberg ion using stimulated Raman adiabatic passage was demonstrated in \cite{Higgins2017b}, this work is the first to demonstrate Rabi oscillations between a LLE state and a Rydberg state.
Such coherent excitation is required for a two-qubit gate using Rydberg-blockade \cite{Jaksch2000} as was realized with neutral atoms \cite{Isenhower2010}.

In summary we use a highly-polarizable ion to investigate the influence of the Stark effect on an ion's trapping potential.
This work provides a solid experimental basis for further studies which rely on the altered trapping potential of Rydberg ions \cite{Li2012, Li2013, Nath2015} and for pursuing quantum information processing with Rydberg ions \cite{Mueller2008}.
In addition we propose using the altered trapping potential of Rydberg ions to simulate cosmological particle creation \cite{Schuetzhold2007, Wittemer2019}, to study quantum fluctuations of work \cite{Huber2008, An2014} and to minimize micromotion beyond the state-of-the-art level.

\begin{acknowledgments}
We thank Weibin Li for theory values of Rydberg state polarizabilities and Tanja Mehlst\"aubler for discussions.
This work was supported by the European Research Council under the European Union's Seventh Framework Programme/ERC Grant Agreement No.\ 279508, the Swedish Research Council (Trapped Rydberg Ion Quantum Simulator), the QuantERA ERA-NET Cofund in Quantum Technologies (ERyQSenS), and the Knut \& Alice Wallenberg Foundation (``Photonic Quantum Information'' and WACQT).
\end{acknowledgments}


%

\section{Supplemental material}

\subsection{Lasers used in the experiment and method for detecting Rydberg excitation}
\begin{figure}
\includegraphics[width=\columnwidth]{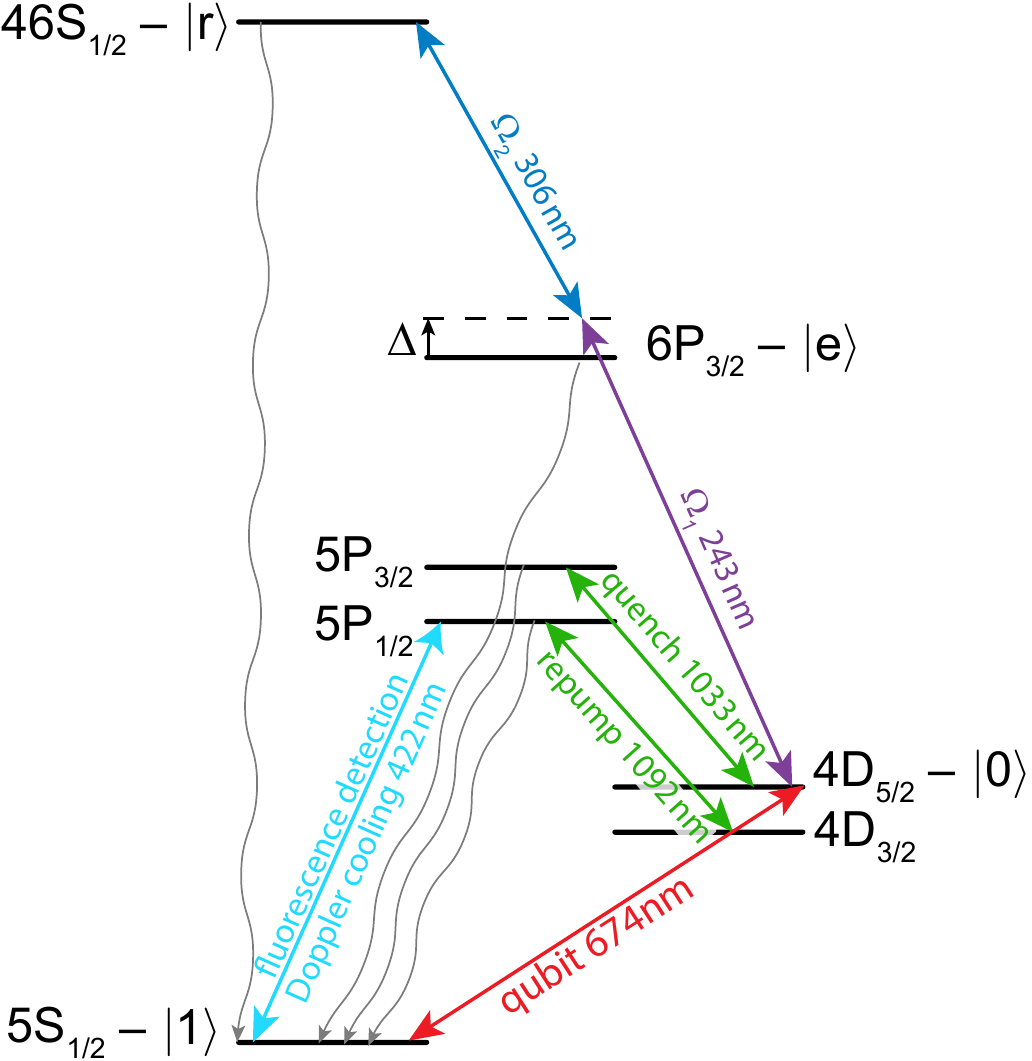}
\caption{
$\mathrm{^{88}Sr^{+}}$ level scheme and lasers used.
\label{supp_fig3}}
\end{figure}

During a single experimental cycle:
The ion is prepared in state $4D_{5/2},m_J=-\tfrac{5}{2} \equiv |0\rangle$ then illuminated by two counterpropagating UV laser beams which drive two-photon Rabi oscillations between the qubit state $|0\rangle$ and the Rydberg state $46S_{1/2},m_J=-\tfrac{1}{2} \equiv |r\rangle$ with detuning $\Delta$ from the intermediate state $6P_{3/2},m_J=-\tfrac{3}{2} \equiv |e\rangle$ (Fig.~\ref{supp_fig3}).
From $|r\rangle$ population decays to $5S_{1/2}$ by multi-channel decay processes in
\textless$\sim20\,\mu s$ \cite{Zhang2016, Safronova2010, Higgins2017a}.
The 422\,nm and 1092\,nm lasers then illuminate the ion; if ion fluorescence is collected on the photo-multiplier tube then the Rydberg-excitation step was successful.

The two UV laser beams counterpropagate along the trap axis.
A magnetic field (0.35~mT) defines the quantization axis along the trap axis.
The two UV lasers are circularly polarized and drive only $\sigma^+$ transitions.

\subsection{Phonon-number changing transitions due to momentum kicks on absorption of photons}
The momentum kicks imparted to the ion during Rydberg excitation by the counterpropagating laser beams largely cancel: the effective Lamb-Dicke parameter is $\eta = 0.045$ with regards the axial mode when $\omega_z = 2 \pi \times 800\,\mathrm{kHz}$ (typical value used).
These momentum kicks do not drive phonon-number changing transitions of the radial modes and do not contribute to the spectra in Fig.~\ref{fig3}(d-f).

\subsection{Relationship between \texorpdfstring{$A$, $B$, $\epsilon$ and $\omega_{x,y,z}$}{alpha, beta, epsilon and the secular frequencies}}

The secular motional frequencies $\omega_{x,y,z}$ are related to the electric field gradients $A$ and $B$ and the radial asymmetry parameter $\epsilon$ by \cite{Li2012}
\begin{align}
A &= \frac{M \Omega}{2 e}\sqrt{\omega_x^2+\omega_y^2+\omega_z^2} \\
B &= \frac{M \omega_z^2}{4 e} \\
\epsilon &= \frac{\omega_y^2 - \omega_x^2}{\omega_z^2}
\end{align}
or conversely
\begin{align}
\omega_x^2 &= \frac{2 e^2 A^2}{M^2 \Omega^2} - \frac{2 e B (1+\epsilon)}{M} \\
\omega_y^2 &= \frac{2 e^2 A^2}{M^2 \Omega^2} - \frac{2 e B (1-\epsilon)}{M} \\
\omega_z^2 &= \frac{4 e B}{M}
\end{align}
where $M$ is the mass of the $\mathrm{^{88}Sr^+}$ ion, $\Omega=2\pi\times18.1\,\mathrm{MHz}$ is the trap drive frequency, $e$ is the charge of the ion.

$\omega_{x,y,z}$ are determined by measuring sidebands on the $5S_{1/2}\leftrightarrow4D_{5/2}$ transition and used to calculate $A$, $B$ and $\epsilon$.

Typical values of trapping parameters used are $\omega_x = 2 \pi \times 1.76\,\mathrm{MHz}$, $\omega_y = 2 \pi \times 1.70\,\mathrm{MHz}$, $\omega_z = 2 \pi \times 0.87\,\mathrm{MHz}$, $A=8.5\times 10^8\,\mathrm{V m^{-2}}$, $B=6.8\times 10^6\,\mathrm{V m^{-2}}$, $\epsilon=-0.26$.

\subsection{Derivation of quadratic Stark shift in Eq.~(\ref{eq_additional_harmonic_potential}) including contributions from intrinsic and excess micromotion, and the effects on the other equations}
The motion of the ion in the radial directions is found by solving the Mathieu equation; the first-order solution of which is given by \cite{Berkeland1998}
\begin{equation}
u_i(t) = (u_{0i} + u_{1i} \cos{\omega_i t})(1 + \frac{q_i}{2} \cos{\Omega t}) \label{eq_ion_motion_using_Berkeland}
\end{equation}
where $i \in \{x,y\}$, $u_{0i}$ describes the equilibrium position in the $i$ direction relative to the oscillating electric field null (in the main text $x_\mathrm{eq}$ and $y_\mathrm{eq}$ are used), $u_{1i}$ is the amplitude of secular motion, and the parameter $q_i$ is given by
\begin{equation*}
q_x = -q_y = \frac{4 e A}{M \Omega^2}
\end{equation*}
In our experiment $q\approx0.29$; linear Paul traps are typically operated with $q<0.5$ \cite{Leibfried2003}.
Intrinsic micromotion has amplitude $\frac{1}{2} u_{1i} q_i$, while excess micromotion has amplitude $\frac{1}{2} u_{0i} q_i$.

The squared electric field experienced by an ion during its motion is
\begin{align}
\begin{split}
\mathcal{E}(t)^2 &\approx \sum_{i} \sum_{j} \Big( \left( \nabla_j \mathcal{E}_i(t) \right) u_j(t) \Big)^2 \\
&\approx \sum_i \left( 2 A \cos{\Omega t} \right)^2 (u_{0i} + u_{1i} \cos{\omega_i t})^2 (1 + \frac{q_i}{2} \cos{\Omega t})^2 \\
&\approx \sum_i 2 A^2 (u_{0i} + u_{1i} \cos{\omega_i t})^2 \left( 1 + \frac{3 q_i^2}{16} \right) \label{eq_new_electric_field}
\end{split}
\end{align}
in the last line the high-frequency $\Omega$ terms were time-averaged.

A polarizable ion experiences a quadratic Stark shift, and so the potential energy of the ion is changed
\begin{equation}
\Delta E_{\mathrm{pot}}(t) \approx -\alpha A^2 \sum_i (u_{0i} + u_{1i} \cos{\omega_i t})^2 \left( 1 + \frac{3 q_i^2}{16} \right) \label{eq_additional_potential_2}
\end{equation}

The first component of Eq.~(\ref{eq_ion_motion_using_Berkeland}) describing the radial motion can be interpreted as motion in the harmonic oscillator (i.e. secular motion)
\begin{equation}
U = \frac{1}{2} M \left[ \omega_x^2 \left( x - x_\mathrm{eq} \right)^2  +\omega_y^2 \left( y - y_\mathrm{eq} \right)^2 \right]
\end{equation}
where
\begin{align}
\begin{split}
x &= u_{0x} + u_{1x} \cos{\omega_x t} \\
y &= u_{0y} + u_{1y} \cos{\omega_y t} \\
x_\mathrm{eq} &= u_{0x} \\
y_\mathrm{eq} &= u_{0y} \label{eq_parameterization}
\end{split}
\end{align}

Substituting Eq.~(\ref{eq_parameterization}) into Eq.~(\ref{eq_additional_potential_2}) shows the change in potential energy due to the Stark shift can be interpreted as resulting from an additional harmonic potential
\begin{equation}
\Delta U = - \alpha A^2 \left( x^2 \left( 1 + \frac{3q_x^2}{16} \right) + y^2 \left( 1 + \frac{3q_y^2}{16} \right) \right) \label{eq_additional_potential_3}
\end{equation}
This equation includes a factor of $(1+\frac{3 q^2}{16}) \approx 1.02$ which was missing from Eq.~(\ref{eq_additional_harmonic_potential}).

Using Eq.~(\ref{eq_additional_potential_3}) instead of Eq.~(\ref{eq_additional_harmonic_potential}) to derive the formulae in the paper means Eqs.~(\ref{eq_omega_prime}), (\ref{eq_Ryd_eq_pos}) and (\ref{eq_energy_shift}) obtain an additional factor of $(1+\frac{3 q^2}{16})$.
Eq.~(\ref{eq_omega_prime}) becomes
\begin{align}
\begin{split}
\omega_x' &\approx \sqrt{ \omega_x^2 - \frac{2 \alpha A^2}{M} \left(1+\frac{3 q_x^2}{16} \right) } \\
\omega_y' &\approx \sqrt{ \omega_y^2 - \frac{2 \alpha A^2}{M} \left(1+\frac{3 q_y^2}{16} \right) } \\
\omega_z' &\approx \omega_z
\end{split}
\end{align}
Eq.~(\ref{eq_Ryd_eq_pos}) becomes
\begin{align}
\begin{split}
x'_{\mathrm{eq}} &= x_{\mathrm{eq}} \left[ 1 - \frac{2 \alpha A^2}{M \omega_x^2} \left(1+\frac{3 q_x^2}{16} \right) \right]^{-1} \\
y'_{\mathrm{eq}} &= y_{\mathrm{eq}} \left[ 1 - \frac{2 \alpha A^2}{M \omega_y^2} \left(1+\frac{3 q_y^2}{16} \right) \right]^{-1}
\end{split}
\end{align}
and Eq.~(\ref{eq_energy_shift}) becomes
\begin{align}
\begin{split}
\delta &= \tfrac{1}{2} M \left( \omega_x^2 x_{\mathrm{eq}}^2 +\omega_y^2 y_{\mathrm{eq}}^2 - {\omega_x'}^2 {x'_{\mathrm{eq}}}^2 - {\omega_y'}^2 {y'_{\mathrm{eq}}}^2 \right)
\\
&\approx - \alpha A^2 \left[ x_{\mathrm{eq}}^2 \left(1+\frac{3q_x}{16}\right) + y_{\mathrm{eq}}^2 \left(1+\frac{3q_y}{16}\right) \right] \\
& \:\:\:\:\:\:\:\:\:= - \tfrac{1}{2} \alpha \langle \mathcal{E}\left( \vec{r}_{\mathrm{eq}}, t \right)^2 \rangle \label{eq_new_energy_shift}
\end{split}
\end{align}
Note that as before the approximation in Eq.~(\ref{eq_new_energy_shift}) reveals $\delta$ is a quadratic Stark shift which results from the electric field at the equilibrium position $\vec{r}_{\mathrm{eq}}$ acting on the polarizable Rydberg ion.
This approximation is valid provided $|\alpha A^2 (1+\frac{3q}{16}) | \ll \frac{1}{2} M \omega_{x,y}^2$ such that the Stark shift $\Delta U$ can be treated as a perturbation.
The final equality uses Eq.~(\ref{eq_parameterization}) and Eq.~(\ref{eq_new_electric_field}).

\subsection{Relating an offset voltage on a micromotion compensation electrode and \texorpdfstring{$\langle \mathcal{E}(\vec{r}_\mathrm{eq},t)^2 \rangle$}{the mean-squared electric field at the ion equilibrium position} -- for rescaling of the x-axis in Fig.~\ref{fig2}(b) and analysis of data in Fig.~\ref{fig3}}
First the magnification of the camera image ($19.2 \pm 0.2$) is determined as follows:
The axial trapping frequency $\omega_z$ is found from measuring sidebands on the $5S_{1/2} \leftrightarrow 4D_{5/2}$ transition.
Two ions are trapped and imaged, the magnification of the camera image is found by relating the separation between the ions in the image and the actual separation between the ions \cite{James1998}.

In the second step a single trapped ion is used: When the voltage on one of the pairs of micromotion compensation electrodes is changed, the ion moves in a radial direction which is approximately perpendicular to the plane of the camera image.
When the voltage on the other pair of micromotion compensation electrodes is changed, the ion moves in a radial direction which is approximately in the plane of the camera image -- this is the voltage which is varied to obtain the data in Fig.~\ref{fig2}(b) and Fig.~\ref{fig3}.
The voltage change is related to the change in ion equilibrium position $\vec{r}_\mathrm{eq}$ using the camera magnification; finally $\langle \mathcal{E}(\vec{r}_\mathrm{eq},t)^2 \rangle$ is found using Eq.~(\ref{eq_electric_pot_w_mm}).

\subsection{Model used in Fig.~\ref{fig3}(d-f)}
The experiment in Fig.~\ref{fig3} is modeled as an optical pumping process from $4D_{5/2}$ to $5S_{1/2}$ and $4D_{3/2}$ via scattering off the Rydberg state $46S_{1/2}$ and the intermediate state $6P_{3/2}$.
Thus, the population remaining in $4D_{5/2}$:
\begin{equation}
\text{population in }4D_{5/2} = 1 - e^{- \mathcal{R}T} \label{eq_optical_pumping}
\end{equation}
where $\mathcal{R}$ is the scattering rate and $T$ is the time for which the Rydberg-excitation lasers are turned on.

When the electric field nulls overlap and phonon-number changing transitions are not excited strongly, then $\mathcal{R}$ is well-described by a single Lorentzian function
\begin{equation}
\mathcal{R} = \frac{\Omega^2\Gamma}{\Gamma^2+4(\omega_{\mathrm{2photon}}-\omega_c)^2 } + \mathcal{R}_{\mathrm{bg}}
\end{equation}
where $\Omega$ is the effective two-photon Rabi frequency, $\Gamma$ is the linewidth, $\omega_{\mathrm{2photon}}$ is the sum of the Rydberg-excitation laser frequencies, $\omega_c$ is the frequency of the phonon-number preserving transition, $\mathcal{R}_{\mathrm{bg}}$ accounts for off-resonant scattering off the intermediate $6P_{3/2}$ state which gives rise to a constant background signal.
$\omega_c$ depends on the number of radial phonons
\begin{equation*}
\omega_c = \omega_0 + \left( n_x+\frac{1}{2} \right) (\omega_x'-\omega_x) + \left( n_y+\frac{1}{2} \right) (\omega_y'-\omega_y)
\end{equation*}
according to Eq.~(\ref{eq_phonon_number_dependent_energy_shift}), here $\omega_0$ is the $4D_{5/2} \leftrightarrow 46S_{1/2}$ resonance frequency for an ion in free space.

When the electric field nulls do not overlap phonon-number changing transitions may be strongly driven.
$\mathcal{R}$ is then described by a sum of Lorentzian functions
\begin{align}
\begin{split}
\mathcal{R} &= \sum_j \sum_k {\mathcal{I}_{x,j}}^2 {\mathcal{I}_{y,k}}^2 \frac{\Omega^2 \Gamma}{\Gamma^2+4(\omega_{\mathrm{2photon}}-\omega_c - j\omega_x' - k\omega_y')^2} \\
&\:\:\:\:+ \mathcal{R}_{\mathrm{bg}} \label{eq_R_longer}
\end{split}
\end{align}
where the contributions from the different Lorentzian functions are described by the Franck-Condon factors
\begin{align}
\begin{split}
\mathcal{I}_{x,j} = \left| \int \psi_{n_x+j}^*(x'_\mathrm{eq}, \omega_x') \psi_{n_x}(x_\mathrm{eq}, \omega_x) dx \right| \\
\mathcal{I}_{y,k} = \left| \int \psi_{n_y+k}^*(y'_\mathrm{eq}, \omega_y') \psi_{n_y}(y_\mathrm{eq}, \omega_y) dy \right|
\end{split}
\end{align}
where $\psi_n(x,\omega)$ is the $n^{\mathrm{th}}$ eigenstate of the one-dimensional quantum harmonic oscillator centered at $x$ with frequency $\omega$.
Franck-Condon factors are shown in Fig.~\ref{fig3}(a-c).
In Eq.~(\ref{eq_R_longer}) the phonon-number preserving transition frequency $\omega_c$ includes the quadratic Stark shift $\delta$ resulting from $\langle \mathcal{E}(\vec{r}_\mathrm{eq},t)^2 \rangle$ described in Eq.~(\ref{eq_energy_shift})
\begin{equation*}
\omega_c = \omega_0 + \left( n_x+\frac{1}{2} \right) (\omega_x'-\omega_x) + \left( n_y+\frac{1}{2} \right) (\omega_y'-\omega_y) + \frac{\delta}{\hbar}
\end{equation*}
In Eq.~(\ref{eq_R_longer}) the Lorentzian function parameterized by $\{j,k\}$ describes the resonance line which corresponds to the transition in which the number of phonons in the x-mode increases by $j$ and the number of phonons in the y-mode increases by $k$.
Note that $j$ and $k$ can take negative values, with the constraint that $n_x + j \geq 0$ and $n_y + k \geq 0$.
Also note that
\begin{equation}
\sum_j {\mathcal{I}_{x,j}}^2 = \sum_k {\mathcal{I}_{y,k}}^2 = 1
\end{equation}

Because $\omega_x' \approx \omega_y'$ we simplify Eq.~(\ref{eq_R_longer})
\begin{align}
\begin{split}
\mathcal{R} &= \sum_r \left( \sum_s {\mathcal{I}_{x,r-s}}^2 {\mathcal{I}_{y,s}}^2 \right) \frac{\Omega^2 \Gamma}{\Gamma^2+4(\omega_{\mathrm{2photon}}-\omega_c - r\omega')^2} \\
&\:\:\:\:+ \mathcal{R}_{\mathrm{bg}} \label{eq_R_shorter}
\end{split}
\end{align}
where $\omega' = \tfrac{1}{2} ( \omega_x' + \omega_y' )$.
Now the $r^{\mathrm{th}}$ term describes the resonance line corresponding to the set of transitions in which the number of radial phonons increases by $r$.

Eq.~(\ref{eq_optical_pumping}) and Eq.~(\ref{eq_R_shorter}) describe the model used.

\subsection{Experimental constraints of model parameters in Fig.~\ref{fig3}(d-f)}
In Fig.~\ref{fig3}(d-f) there are the Rydberg-excitation spectra for an ion prepared with $n_x=20, n_y\approx0$.
The parameters of the model [Eq.~(\ref{eq_R_shorter})] are constrained as follows:

$\omega_x$ and $\omega_y$ are found using spectroscopy on the $5S_{1/2} \leftrightarrow 4D_{5/2}$ transition.
$\omega_x'$ and $\omega_y'$ are found using Eq.~(\ref{eq_omega_prime}).
We determine $\omega_x' / \omega_x = (0.9740 \pm 0.0004)$ for the measurements in Fig.~\ref{fig3}(d-f); and so the calculations in Fig.~\ref{fig3}(a-c) used $\omega_x' / \omega_x = 0.974$.
The mismatch of phonon states in Fig.~\ref{fig3}(b,c) and the phonon-number changing transitions of Fig.~\ref{fig3}(e,f) are mostly due to the difference in equilibrium positions; the difference in trap stiffness ($\omega_x' \neq \omega_x$) also contributes to the mismatch and to phonon-number changing transitions, as seen in Fig.~\ref{fig3}(a).

$x_\mathrm{eq}$, $x'_\mathrm{eq}$, $y_\mathrm{eq}$ and $y'_\mathrm{eq}$ are estimated by using the method above to relate the offset voltage on the micromotion compensation electrode to the equilibrium position $\vec{r}_\mathrm{eq}$, then using Eq.~(\ref{eq_Ryd_eq_pos}) to estimate $\primedvector{r}'_\mathrm{eq}$.

\begin{figure*}[ht]
\centering
\includegraphics[width=\textwidth]{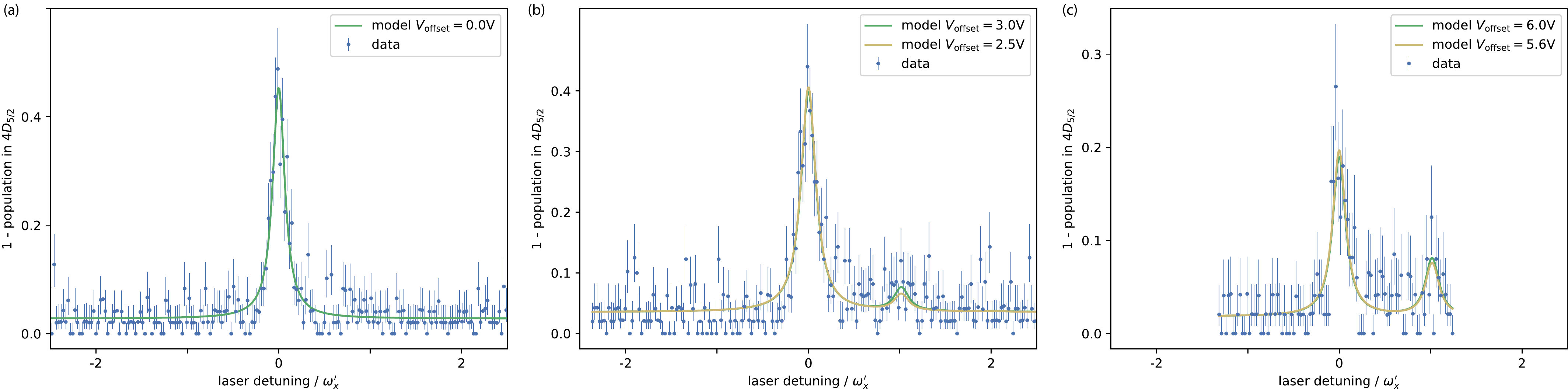}
\caption{Resonance scans with the same settings (within the uncertainties of the experimental parameters) as in Fig.~\ref{fig3}(d-f), but with $n_x\approx0$ and $n_y\approx0$. The data here is fit using the model in Eq.~(\ref{eq_R_shorter}) with $\Omega$, $\Gamma$, $\omega_c$ and $\mathcal{R}_{\mathrm{bg}}$ as free parameters.
The fit results are used to constrain the parameters in Fig.~\ref{fig3}(d-f).
Error bars indicate projection noise (68\% confidence interval).
}
\label{supp_fig}
\end{figure*}

$\Omega$, $\Gamma$, $\omega_c$ and $\mathcal{R}_{\mathrm{bg}}$ are estimated by measuring Rydberg-excitation spectra for an ion prepared with $n_x \approx 0, n_y \approx 0$ and fitting the experimental data using the model [Eq.~(\ref{eq_R_shorter})].
These excitation spectra are shown in Fig.~\ref{supp_fig}(a-c) -- the experimental parameters are similar to the parameters used in Fig.~\ref{fig3}(d-f).
The value of $\omega_c$ determined from Fig.~\ref{supp_fig}(a-c) is corrected by $20(\omega_x'-\omega_x) = 2\pi\times838\,\mathrm{kHz}$ to account for the difference in the contribution of $(n_x+\tfrac{1}{2})(\omega_x'-\omega_x)$ to $\omega_c$ between the data in Fig.~\ref{supp_fig}(a-c) and the data in Fig.~\ref{fig3}(d-f).

\subsection{Limits of micromotion compensation using Rydberg ion spectroscopy and theory values for Rydberg state polarizabilities}
\begin{figure}[ht!]
\centering
\includegraphics[width=\columnwidth]{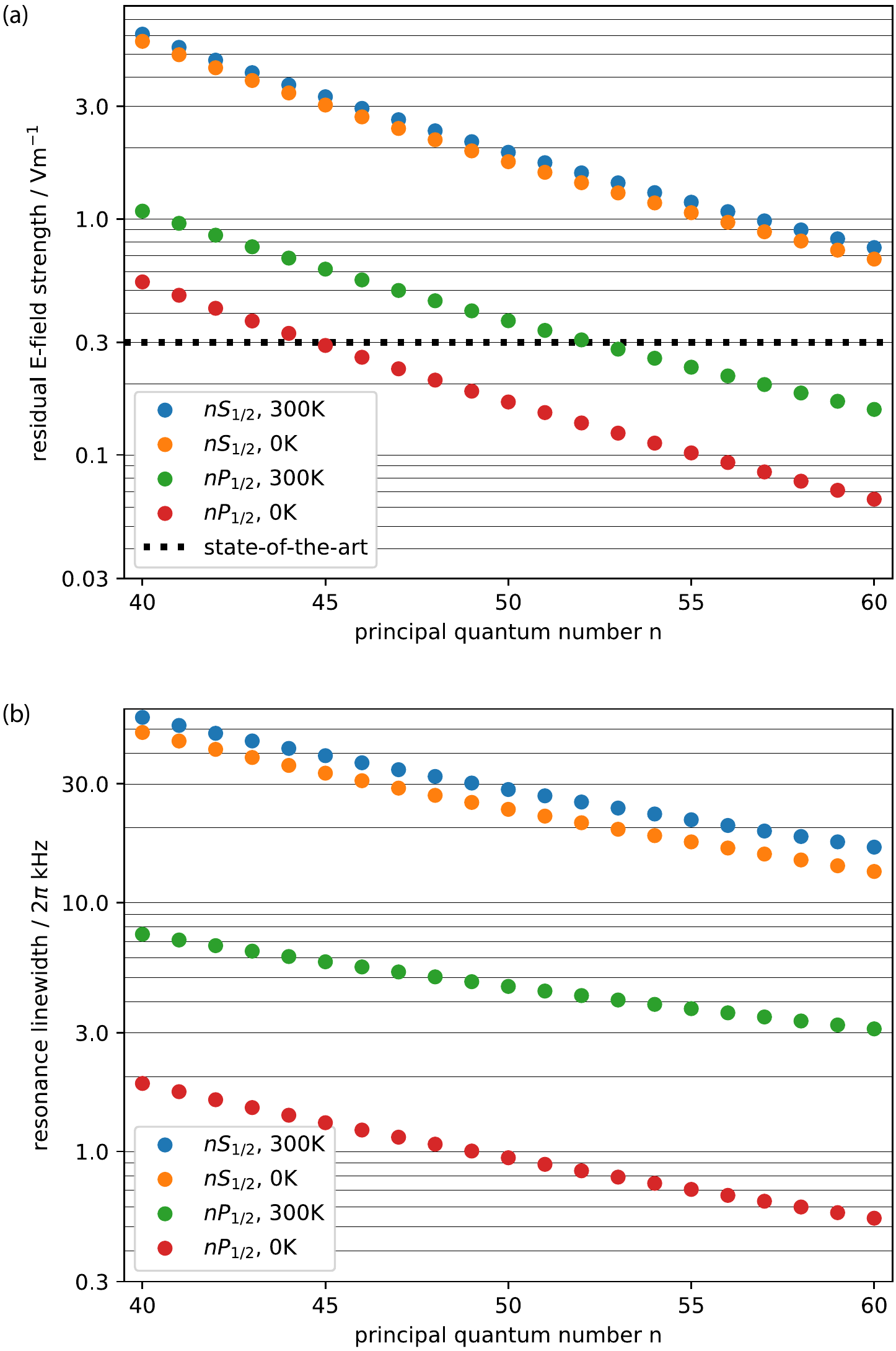}
\caption{Using Rydberg spectroscopy for micromotion minimalization.
(a) Estimated residual rf electric field strength that could be reached using different Rydberg states, estimated using Eq.~(\ref{eq_MM_lim}) with resolved resonance shifts 10 times lower than the resonance linewidth ($F=0.1$).
Rydberg $S_{1/2}$ and $P_{1/2}$ states are compared, in environments at different temperatures (0\,K environment uses natural linewidth).
(b) Resonance linewidths.}
\label{supp_fig_MM}
\end{figure}

If one can determine the center of a Rydberg resonance line with width $\Delta \nu$ to precision $F \times \Delta \nu$, then one can resolve Stark shifts
\begin{equation}
F \times h \Delta \nu = \frac{1}{2} \alpha \mathcal{E}_{\mathrm{res}}^2 \label{eq_MM_lim}
\end{equation}
With $F=0.1$ we estimate the lowest residual rf electric field strength $\mathcal{E}_{\mathrm{res}}$ that we can resolve using different Rydberg states, the results are shown in Fig.~\ref{supp_fig_MM}(a).
Rydberg $P_{1/2}$-states have higher polarizabilities and lower resonance linewidths than Rydberg $S_{1/2}$-states, and thus Rydberg $P_{1/2}$ states may be used to reduce the residual electric field strength further.
Using Rydberg state $53P_{1/2}$ at room temperature the residual rf electric field strength may be reduced to the state-of-the-art level \cite{Harter2013}.
Rydberg $P_{1/2}$ states are excited from state $|0\rangle$ by a three-photon transition, using the two UV laser fields as well as a microwave field.

It is important to consider the time required to resolve the resonance shifts, and so in Fig.~\ref{supp_fig_MM}(b) the resonance linewidths are plotted.
In the room temperature environment, to resolve resonance shifts $\sim 10$ times lower than the resonance linewidths, millisecond laser pulses are required.
Such laser pulses would not limit the time required to conduct micromotion compensation any more than other steps of an experimental sequence (such as sideband cooling or electron shelving).

\subsection{Preparation of phonon number states}
The ion is prepared in $5S_{1/2},m_J=-\frac{1}{2},n_x\approx0,n_y\approx0$ using resolved sideband cooling and optical pumping.

To add $n_x$ phonons in the x-radial mode the following cycle is repeated $n_x$ times:
\begin{enumerate}
\item A $\pi$-pulse is applied on the blue radial-x sideband of the $5S_{1/2},m_J=-\frac{1}{2} \leftrightarrow 4D_{5/2},m_J=-\frac{5}{2}$ transition.
\item A pulse of 422\,nm and 1092\,nm light is applied, the photomultiplier tube collects fluorescence at 422\,nm if the ion was in state $5S_{1/2}$ which signifies that the previous step was unsuccessful. Such instances are removed when the data is analyzed.
\item A $\pi$-pulse is applied on the $5S_{1/2},m_J=-\frac{1}{2} \leftrightarrow 4D_{5/2},m_J=-\frac{5}{2}$ transition, which returns population to $5S_{1/2},m_J=-\frac{1}{2}$.
\item To ensure population is returned to $5S_{1/2},m_J=-\frac{1}{2}$, a short ($\sim$1\,$\mu$s) 1033\,nm pulse is applied.
\end{enumerate}
The $\pi$-time in step 1 is varied to account for the $\sqrt{n+1}$ dependence of the blue sideband Rabi frequency.

Our method is similar to the method reported in \cite{Meekhof1996}; our method has the advantage that post-selection increases the fidelity of the preparation.

\begin{figure*}
\includegraphics[width=\textwidth]{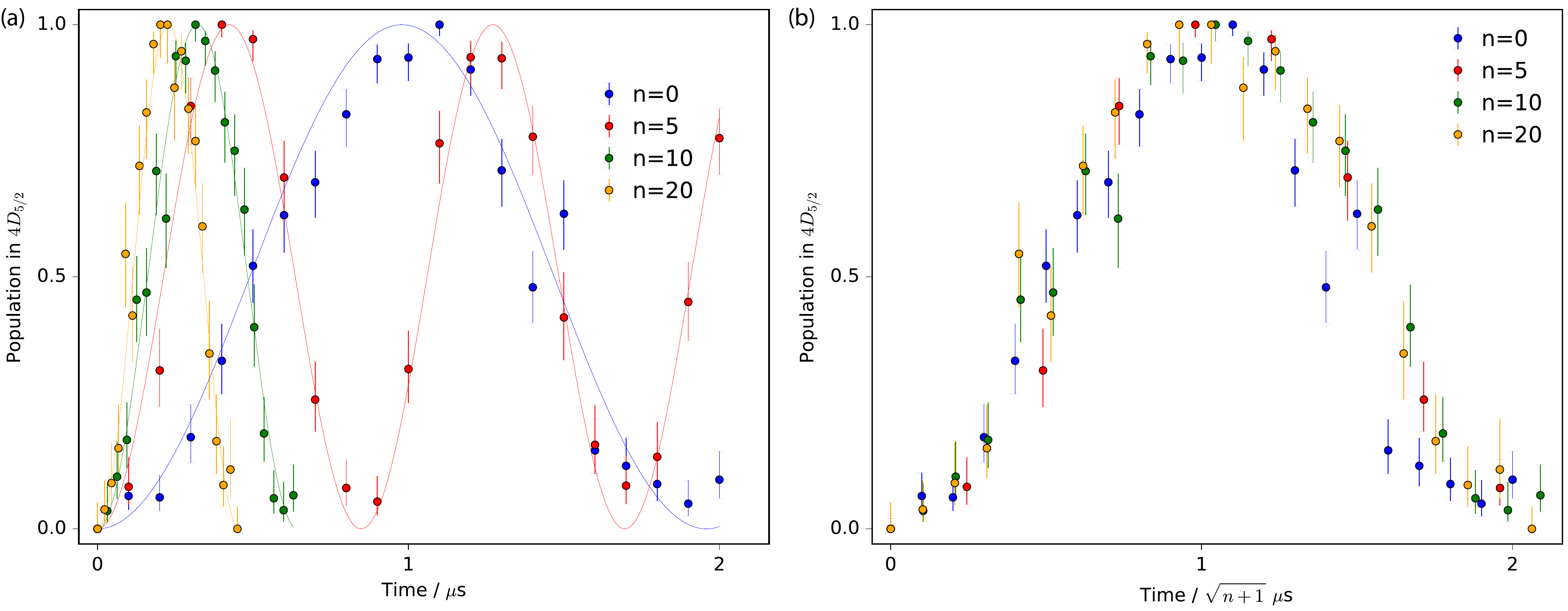}
\caption{
Test of phonon-number state preparation.
The ion is prepared in $5S_{1/2},m_J=-\frac{1}{2}$ with $n_x\approx0$ and $n_y=n$.
Rabi oscillations on the blue radial-y sideband of $5S_{1/2},m_J=-\frac{1}{2} \leftrightarrow 4D_{5/2},m_J=-\frac{5}{2}$ are driven.
High-contrast oscillations are observed and the Rabi frequency scales with $\sqrt{n+1}$ -- this shows we can reliably prepare number states.
In both plots the dots indicate experimental data.
The solid lines in (a) are sinusoidal fits to guide the eye.
Plot (b) has the same datasets with the time axis rescaled for each dataset, to make the $\sqrt{n+1}$-dependence more clear.
Error bars indicate quantum projection noise (68\% confidence interval).
\label{supp_fig2}}
\end{figure*}

To check the preparation of phonon number states we prepare the ion with $n$ phonons in the y-mode, then drive Rabi oscillations on the $5S_{1/2},m_J=-\frac{1}{2} \leftrightarrow 4D_{5/2},m_J=-\frac{5}{2}$ blue sideband transition.
The results are shown in Fig.~\ref{supp_fig2}.
High contrast Rabi oscillations are observed and the Rabi frequency scales as $\sqrt{n+1}$ -- this confirms phonon Fock states are reliably prepared.

\subsection{Experimental constraints of model parameters in Fig.~\ref{fig4}}
The 243\,nm laser detuning is determined by resonantly driving the $|0\rangle \leftrightarrow |e\rangle$ transition, then detuning the laser using an AOM.
The 243\,nm laser Rabi frequency is found from the scattering rate off the $|0\rangle \leftrightarrow |e\rangle$ transition when this laser is detuned and the 306\,nm laser is turned off.
The 306\,nm laser Rabi frequency and detuning is determined using an Autler-Townes splitting as described in the supplemental material of \cite{Higgins2017b}.
The two-photon detuning is determined from the 243\,nm laser detuning and the 306\,nm laser detuning.

The number of phonons in the radial modes after sideband cooling and after Doppler cooling is estimated by comparing the Rabi frequencies of the carrier transition, blue sideband transitions and red sideband transitions of the $5S_{1/2},m_J=-\tfrac{1}{2} \leftrightarrow 4D_{5/2},m_J=-\tfrac{5}{2}$ transition.

\subsection{Simulation results in Fig.~\ref{fig4}}

The Lindblad master equation is numerically solved for the four-level system \{$|0\rangle$,$|e\rangle$,$|r\rangle$,$5S_{1/2}$\} using the open source python framework QuTiP \cite{Johansson2013}.
Only experimental parameters are used.

The two-photon detuning from the Rydberg resonance depends on the number of radial phonons $\Delta_{\text{2-photon}} = n_x \hbar \Delta\omega_x + n_y \hbar \Delta\omega_y$.

To include the effect of phonon number the simulation is repeated with different $\Delta_{\text{2-photon}}$, which accounts for different $n_x, n_y$, and the results are added together with weights which account for the population with phonon numbers $n_x, n_y$.

An alternative approach would be to include additional phononic dimensions in the simulation, however this slows down the simulation making it intractable on a desktop PC.

Since the experimental parameters contain uncertainties, they are randomly sampled and the simulation is repeated 100 times with different parameter values.
The results are analyzed as follows:
at each time step the populations in $|0\rangle$ from the 100 simulations are ordered.
The 16\textsuperscript{th} and the 84\textsuperscript{th} highest values enclose 68\% (one standard deviation) of the values returned by the simulations.
Thus, the area enclosed by the 16\textsuperscript{th} and 84\textsuperscript{th} highest values at each time step is used as the 68\% confidence interval of the simulation results.

As well as increasing $\Delta_{\text{2-photon}}$, the change in trapping potential between $|0\rangle$ and $|r\rangle$ causes phonon-number changing transitions to occur during Rydberg excitation.
The strength of a phonon-number changing transition is described by a Franck-Condon factor.
To check the effect of such transitions, additional simulations were carried out which included phononic dimensions.
Because we believed excess micromotion was well-minimized, we take $\vec{r}_\mathrm{eq} = \primedvector{r}'_\mathrm{eq}$, and the phonon-number changing transitions then result from the difference between $\omega_{x,y}$ and $\omega_{x,y}'$.
The resultant phonon-number changing transitions cause a decrease in Rabi oscillation contrast, though the decrease is small enough to be neglected.

\end{document}